\documentclass[onecolumn,
 aip,
 amsmath,amssymb,
 reprint,%
]{revtex4-1}

\usepackage{graphicx}
\usepackage{dcolumn}
\usepackage{bm}

\usepackage[utf8]{inputenc}
\usepackage[T1]{fontenc}
\usepackage{mathptmx}
\usepackage{etoolbox}
\usepackage{hyperref}

\makeatletter
\def\@email#1#2{%
 \endgroup
 \patchcmd{\titleblock@produce}
  {\frontmatter@RRAPformat}
  {\frontmatter@RRAPformat{\produce@RRAP{*#1\href{mailto:#2}{#2}}}\frontmatter@RRAPformat}
  {}{}
}%
\makeatother
\begin{document}

\preprint{AIP/123-QED}

\title[Orbits with Video Games]{Using Video Games to Teach Kepler's Laws and Orbital Dynamics}

\author{Brian DiGiorgio Zanger}
\affiliation{Physics and Astronomy Department, Swarthmore College, 500 College Avenue, Swarthmore, PA 19081, USA}
\email{bzanger1@swarthmore.edu}

\date{\today}
\maketitle

\section{Introduction}
Physics instructors often rely on demonstrations when teaching, using real-time examples to appeal to student intuition or tinkering with a physical system to develop a deeper and more natural understanding of a topic. However, some subjects, like Kepler's laws of planetary motion, are impractical to demonstrate at a human scale, and when reality proves difficult, it is often easier to move to simulations and video games. In particular, using video games in the classroom benefits student learning \citep{griffiths02} and allows students to learn practical information that they can transfer to the real world \citep{finkelstein05}. The video game \textit{Kerbal Space Program}\cite{kerbalwebsite} (KSP) is notable for its commitment to accurate simulation of rockets and orbits, allowing players to explore the in-game solar system with their Kerbal astronauts only if they have mastered the laws of orbital dynamics. In my personal experience, I gained a better intuition about orbits during my years playing KSP than I did in graduate school, a common sentiment within the KSP community.\cite{xkcd1356} I use KSP in my college introductory astronomy courses to make Kepler's laws more tactile with demonstrations and projects, developing intuition on advanced aspects of rocket science through experiments that are otherwise impractical at interplanetary scales. 

Past authors have attempted to demonstrate Kepler's laws in the classroom in a number of ways. Some have used rolling objects in funnels to demonstrate Kepler's first and second laws \citep{ogawara18, worner21}, mathematical and physical representations of equal area triangles for Kepler's second law \citep{rodrigues22, stoekel23}, and centripetal force to experimentally derive Kepler's third law \citep{lin93}. There are also freely available online applets (eg. from PhET\citep{phet} and NAAP\citep{naap}) that allow for high-level tinkering in simplified simulations. 

Previous studies have used video games extensively in physics instruction to increase engagement, performing experiments in a novel and engaging environment \citep{heuvel24, nordine11}, using them to clear up physics misconceptions \citep{disessa82}, or exploring the physics of an unknown virtual environment \citep{heuvel16, mohanty11, allain09}. Video games also enable physics teachers to emulate environments beyond what is possible in the classroom, as KSP does with orbital dynamics \citep{blanco19, blanco17}.

KSP, released in 2015, is available from the \textit{Steam} game store for \$40 (often on sale for \$10), is compatible with all modern operating systems, and requires comparatively few graphical and computational resources to run smoothly. It is also available for some game consoles, though the controls are more difficult than on computer, so they are not recommended. The sequel, \textit{Kerbal Space Program 2}, is inferior to the original for practical purposes and should not be used in classrooms. The game has a steep learning curve and is difficult to master due to the inherent restrictions of space flight, but it has infinite depth and maintains an active online community even a decade after its release. Readers can access files containing the save data for the rockets in this paper, all of which are constructed for simple and easy operation and have a sizable margin for error on fuel, at TPT Online at XXXX in the “Supplementary Material” section.

\section{Teaching Kepler's Laws in KSP}
\begin{figure}
    \centering
    \includegraphics[width=\linewidth]{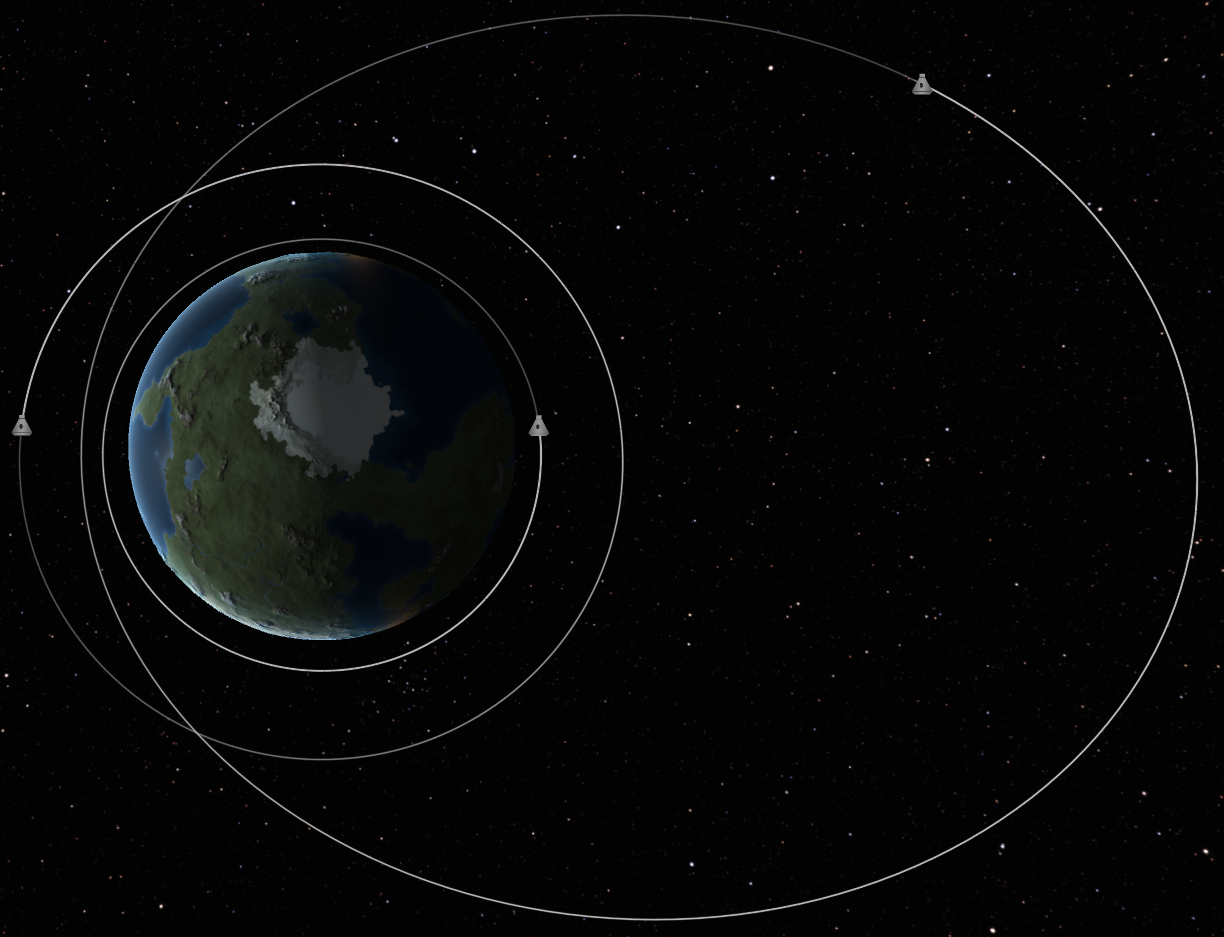}
    \caption{A setup of three orbiting spacecraft} in KSP for showing Kepler's laws. Both circular and elliptical orbits are seen, and two circular orbits at different radii demonstrate different orbital periods.
    \label{kepler}
\end{figure}

The ability to play in an accurate orbital sandbox allows for the intuitive discovery of Kepler's laws in real time and with more direct control than traditional applets. After launching a rocket into orbit on the classroom projector into a game setup like the one in Figure \ref{kepler}, I step through the following problems. I have performed the following demonstrations during a single hour-long lecture period of college-level introductory courses for nonmajors at liberal arts colleges. Other instructors may choose to supplement the KSP demonstration by allowing students to experiment on their own, either with KSP or with other purpose-built teaching simulations like PhET.

\subsection{Kepler's First Law}
\textbf{What shape is the rocket's orbit?} A normal orbit will appear to be circular in map view, but after providing an impulse with the rocket's engine, students can watch as the orbit slowly elongates into an ellipse with increasing eccentricity. With this demonstration and perhaps complementary time with other interactive tools, the class can intuit that all closed orbits must be elliptical with the planet at the focus, and that circular orbits are a special case of elliptical orbits. 

\begin{figure}
    \centering
    \includegraphics[width=\linewidth]{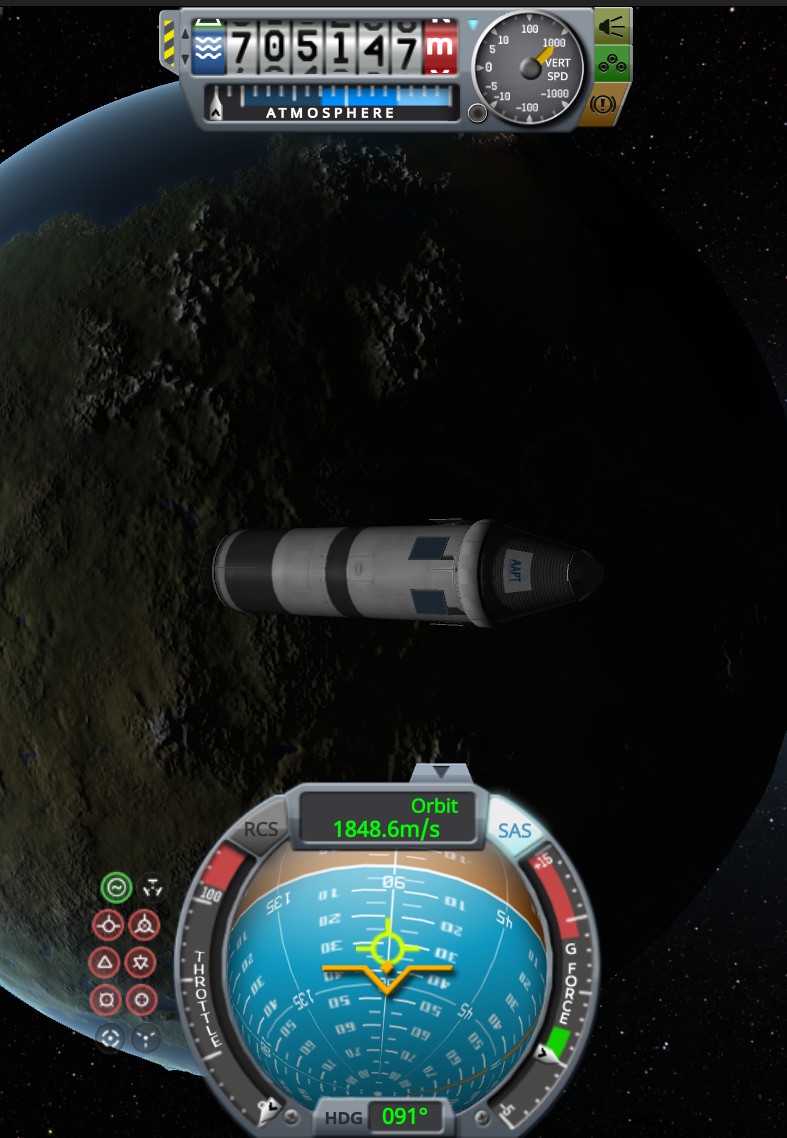}
    \caption{A simple rocket in orbit, showing the altitude (top) and orbital velocity (bottom) needed to understand Kepler's second law.}
    \label{rocket}
\end{figure}

\subsection{Kepler's Second Law}
\textbf{How fast does the rocket go throughout its orbit?} KSP displays a readout of the rocket's speed throughout its orbit (Figure \ref{rocket}), so by increasing the simulation rate, the class can watch as the rocket speed increases at smaller radius and decreases when further away. This change can also be seen in map mode while watching the rocket's position, similar to other online simulations.

\subsection{Kepler's Third Law}
\textbf{How long does the rocket take to complete an orbit?} The class can compare the relative orbital periods between two vessels at different orbital radii and observe that radius is proportional to orbital period, a conclusion that can be expressed more mathematically with other purpose-built applets.

\section{Project: Moon Landing}
\begin{figure}
    \centering
    \includegraphics[width=\linewidth]{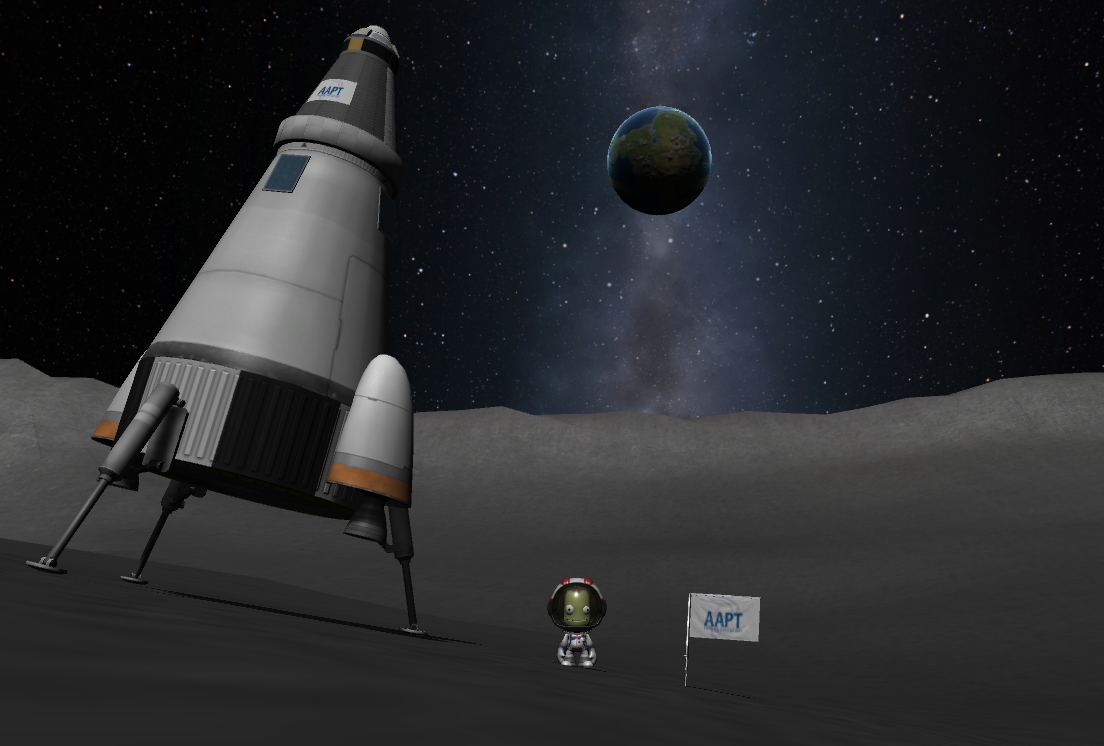}
    \caption{A Kerbal after successfully planting a flag on the M\"un, with their lander, a crater, and the Kerbal home planet also visible.}
    \label{mun}
\end{figure}

After demonstrating the basics of KSP in class, I give students the option to do their final course project on landing a Kerbal on the M\"un (the Kerbal moon) and bringing them back (Figure \ref{mun}). Doing so requires the student to understand and carry out a number of concepts that fall far beyond the typical scope of an introductory astronomy course but are easily contextualized in a practical framework by KSP. I do not formally assess students on these topics beyond their ability to perform the landing itself, but I ask students to submit a written reflection on the physics concepts they learned.

\begin{itemize}
    \item \textbf{Spacecraft construction:} A student must put Newtonian mechanics into practice by building a rocket with sufficient thrust to counteract gravity and fuel to provide impulse. KSP has a modular rocket construction interface to facilitate simple rocket construction.
    \item \textbf{Launch to orbit:} A successful and efficient launch to orbit requires a student to achieve orbital velocity, putting Newton's Cannonball thought experiment into action. KSP shows rocket trajectories in real time, allowing students to instantly see the results of their engine burns.
    \item \textbf{Orbital circularization:} Entering a controlled orbit requires an understanding of elliptical orbit shape, apoapsis, periapsis, and different thrust vectors for adjusting orbits, all of which are conveyed through the KSP interface.
    \item \textbf{Orbital transfer:} Students must pick the correct timing and trajectory to complete the orbital transfer to the M\"un. KSP has a simple system for planning orbital maneuvers in advance, encouraging students to solve these problems through experimentation. 
    \item \textbf{Soft landing:} Though not strictly part of orbital dynamics, landing softly on another body is one of the most common failure points for real interplanetary missions. The low-stakes environment of a video game lets students quickly learn from their inevitable mistakes and crashes.
    \item \textbf{Return and reentry:} Returning a Kerbal safely home and not burning up in the atmosphere on reentry requires additional planning for greater fuel and heating needs. 
\end{itemize}

Though the launch and landing themselves can be performed relatively quickly, learning and performing the skills above is a significant time investment (typically 10+ hours), so I recommend assigning this task as a long-term project or over multiple consecutive lab periods. There are several extensive and useful tutorials that walk through the KSP M\"un landing process on YouTube\cite{kspyoutube} as well as other more general tutorial series on the game, and I encourage students to take full advantage of these resources. In their reflections, students who have completed this project have expressed satisfaction in completing the skill in combination with frustration at the restrictions of physics, and they have demonstrated a deep understanding of their orbital maneuvers when describing their mission.

A full project description and rubric is available at TPT Online at XXXX in the “Supplementary Material” section.

\section{Advanced Activities}
In addition to the basics of Kepler's laws, more advanced questions in orbital dynamics can be answered from experimentation in KSP. Instructors who devote more time to the subject may wish to explore these ideas:

\begin{itemize}
\item \textbf{Which orbits have more energy?) Larger orbits have more energy since the fuel's energy goes into speeding up the rocket and ultimately raising its orbit, showcasing the relationship between orbital radius and total energy.}

\item\textbf{What happens when an elliptical} orbit gets too large? Increasing speed beyond escape velocity shows the shape of parabolic and hyperbolic orbits at greater orbital energy.

\item\textbf{What is the specific relationship between period and semimajor axis?} KSP always shows the exact height above the surface of the Kerbal planet, so by adding on its planetary radius (600 km), the semimajor axis of any orbit can be calculated and plotted to derive $P \propto a^{3/2}$. 

\item\textbf{How did humans go to the Moon?} Two Kerbal spacecraft can be slowly and deliberately brought together by taking advantage of Kepler's third law to maneuver in orbit. This can spur a discussion of Buzz Aldrin's contributions to the theory of docking \citep{aldrin63} and the mission profile of the Apollo moon landings.

\item\textbf{How do spacecraft travel from one planet to another?} KSP simulates an entire solar system, allowing for Hohmann transfer orbits between different planets. These transfers can be calculated using online tools\citep{kspcalcs} or can be done by hand in an advanced course.

\item\textbf{What is the easiest way to gain orbital energy?} KSP accurately simulates the Oberth effect, an unexpected consequence of Newtonian relativity that means that more kinetic energy is gained while burning rocket fuel at small orbital radius rather than at large orbital radius \citep{blanco19}.
\end{itemize}

\section{Conclusion}
\textit{Kerbal Space Program} is a powerful and flexible physics sandbox that faithfully simulates topics that are otherwise difficult to represent tactilely in a physics classroom. It allows students to see orbital mechanics happen in real time, to experiment with complex missions on their own, and to learn from their mistakes in a low-stakes environment. Though the game is difficult, success is built off of a deep conceptual understanding of the orbits and Kepler's laws.

\vspace{10pt}
I would like to acknowledge the developers of KSP for creating a game that has inspired so many, the online KSP community for teaching me to play and learn from this game, and the reviewers who helped to improve this paper. 

\vspace{10pt}
Brian DiGiorgio Zanger is a visiting assistant professor of astronomy at Swarthmore College and is a proud proponent of physics and astronomy in liberal arts settings.

\bibliography{ref}

\end{document}